\documentclass[preprint,tightenlines,aps,floats,nofootinbib,showpacs]{revtex4}

\usepackage[dvips]{color}
\usepackage{slashed}
\usepackage{tabularx}
\usepackage{epsf}
\usepackage{epsfig}
\usepackage{graphicx}

\newcommand{\degree} {^{\circ}}

\newcommand{\GeV} { \textrm{GeV}}
\newcommand{\sGeV}{\textrm{ GeV}}

\newcommand{\sTeV}{\textrm{ TeV}}

\newcommand{\sfb}{\textrm{ fb}}

\newcommand{\sifb}{\textrm{ fb}^{-1}}

\newcommand{\siab}{\textrm{ ab}^{-1}}

\newcommand{\MET}{\slashed{E}_T}

\newcommand{\Lagr}{\mathcal{L}}

\newcommand{\calpha}{\cos\alpha}
\newcommand{\salpha}{\sin\alpha}
\newcommand{\cbeta}{\cos\beta}
\newcommand{\sbeta}{\sin\beta}
\newcommand{\tbeta}{\tan\beta}

\newcommand{\mgluino}{ m_{\tilde{g}} }
\newcommand{\msquark}{ m_{\tilde{q}} }
\newcommand{\msbottom}[1]{ m_{\tilde{b}_#1} }
\newcommand{\mstop}[1]{m_{\tilde{t}_#1}}

\newcolumntype{L}[1]{>{\raggedright\let\newline\\\arraybackslash\hspace{0pt}}m{#1}}
\newcolumntype{C}[1]{>{\centering\let\newline\\\arraybackslash\hspace{0pt}}m{#1}}
\newcolumntype{R}[1]{>{\raggedleft\let\newline\\\arraybackslash\hspace{0pt}}m{#1}}

\preprint{\font\fortssbx=cmssbx10 scaled \magstep2
\hbox to \hsize{
\hskip1.2in 
\hbox{\fortssbx The University of Oklahoma}
\hskip0.2in $\vcenter{
                      \hbox{\bf OUHEP-130923}
                      \hbox{\bf arXiv: [hep-ph]}
                      \hbox{December 2013}}$ }
}

\begin{document}


\begin{picture}(20,20)(0,28)
\includegraphics[scale=0.2]{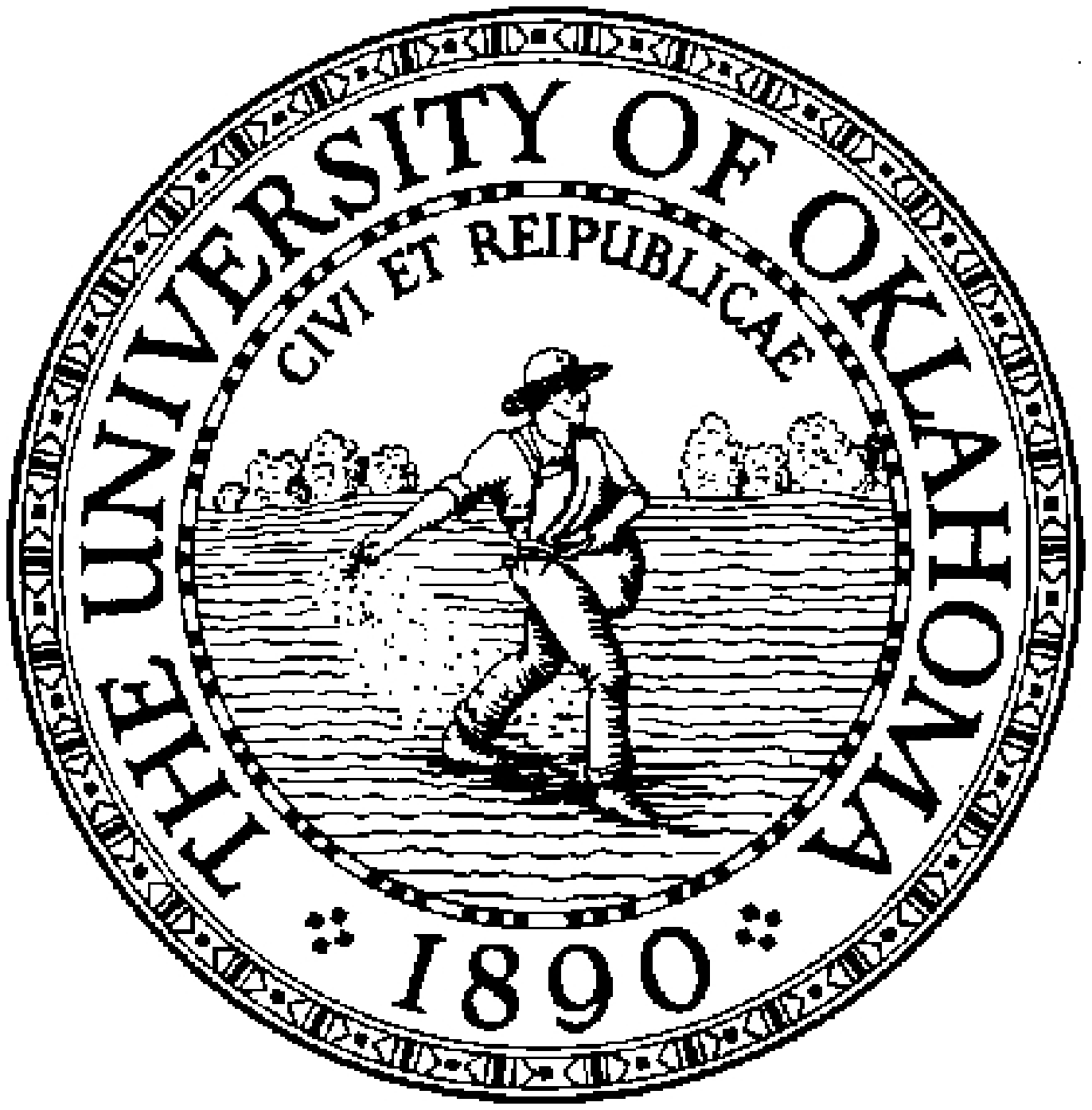}
\end{picture}

\title{\vspace*{0.7in}
Unveiling the MSSM Neutral Higgs Bosons\\with Leptons and a Bottom Quark}

\author{Baris Altunkaynak, Chung Kao and Kesheng Yang}

\affiliation{
Homer L. Dodge Department of Physics and Astronomy, 
University of Oklahoma, Norman, OK 73019, USA 
\vspace*{0.5in}}

\date{December 11, 2013}

\thispagestyle{empty}

\begin{abstract}

We investigate the prospects for the discovery of neutral Higgs bosons 
produced with a bottom quark where the Higgs decays into a pair of 
tau leptons and the taus decay into an electron-muon pair, i.e. 
$bg \to b\phi^0 \to b\tau^+\tau^- \to be^\pm\mu^\mp +\MET$, $ 
\phi^0 = h^0, H^0, A^0$. 
Our study has been done within the framework of the Minimal
Supersymmetric Standard Model. 
We consider the dominant physics backgrounds including the production of 
Drell-Yan processes ($b\tau^+\tau^-$ and $j\tau^+\tau^-, j = q, g$), 
top quark pair ($t\bar{t}$), $tW$ and $jWW$ with realistic acceptance
cuts and efficiencies.
We present $5\sigma$ discovery contours for the neutral Higgs bosons in the 
($M_A,\tan\beta$) plane as well as the region with a favored light
Higgs mass (123 GeV $\le m_h\le$ 129 GeV).
Promising results are found for the CP-odd pseudoscalar ($A^0$) 
and the heavier CP-even scalar ($H^0$) Higgs bosons with masses up to
800 GeV and $\tan\beta \simeq 50$ at the LHC with a center of mass
energy ($\sqrt{s}$) of 14 TeV and an integrated luminosity ($L$) 
of 300 fb$^{-1}$. 
With $\sqrt{s} =$ 14 TeV and $L =$ 3000 fb$^{-1}$, LHC will be able
to discover the Higgs pseudoscalar and the heavier Higgs scalar beyond 
$M_A = 1000$ GeV.

\end{abstract}


\maketitle

\newpage

\section{Introduction}

Recent discovery of the Higgs boson by the ATLAS and the CMS 
experiments~\cite{atlas-higgs-discovery,cms-higgs-discovery} 
has completed the remaining piece of the standard electroweak symmetry
breaking (EWSB) puzzle and has one more time confirmed
the success of the Standard Model (SM). 
Despite its success we know that the Standard Model is not a complete theory and 
there is new physics to be discovered at or beyond the electroweak
scale. After this remarkable achievement, the goal is now to discover signs of
new physics with particles and interactions beyond the Standard Model.

One of the most studied new physics candidate is the supersymmetric
extension of the Standard Model. Supersymmetry (SUSY) is very well motivated
both theoretically and phenomenologically and its realization with minimal 
particle content is called the Minimal Supersymmetric Standard Model (MSSM).
The extensive search for the signs of SUSY and MSSM has so far only 
returned exclusion limits for SUSY particle masses. 
For simplified models, the current limits
are above a TeV for gluinos and first/second generation squarks, and
hundreds of GeV for electroweak 
gauginos~\cite{ATLAS-SUSY-exclusion1,ATLAS-SUSY-exclusion2,CMS-SUSY-exclusion1}.

The MSSM Higgs sector consists of two $SU(2)$ doublets 
$\phi_1$ and $\phi_2$ that couple to fermions with weak isospin 
$t_3 = -1/2$ and $t_3 = +1/2$, respectively \cite{Guide}. 
After spontaneous symmetry breaking, there remain five physical Higgs
bosons: a pair of singly charged Higgs bosons $H^{\pm}$,
two neutral CP-even scalars $H^0$ (heavier) and $h^0$ (lighter),
and a neutral CP-odd pseudoscalar $A^0$.
At the tree level, all properties of the Higgs sector are fixed by 
two parameters that are usually chosen to be the Higgs pseudoscalar 
mass ($m_A$) and the ratio of the vacuum expectation values of 
the two Higgs doublets ($\tan\beta \equiv v_2/v_1$). 
In the decoupling limit~\cite{Gunion:1984yn} with $\tan\beta \agt 10$ 
and $m_A \agt 150$, the light Higgs scalar behaves like 
the SM Higgs boson, while the heavy Higgs scalar ($H^0$) and 
the Higgs pseudoscalar ($A^0$) are almost degenerate 
in mass with dominant decays into $b\bar{b}$ ($\sim 90\%$) and 
$\tau^+\tau^-$ ($\sim 10 \%$) final states.

A supersymmetric light Higgs boson of mass $126 \sGeV$ implies large loop
corrections to the tree level Higgs mass which requires a heavy stop
and/or large trilinear couplings \cite{Baer:2011ab,Akula:2011aa}. 
These large loop corrections also
indicate large fine tuning.  Although low fine tuned MSSM is still a
possibility~\cite{Baer:2013gva}, the available parameter space is
shrinking. Non-observation of  superpartners so far indicate a heavy
SUSY particle 
spectrum~\cite{ATLAS-SUSY-exclusion1,ATLAS-SUSY-exclusion2,CMS-SUSY-exclusion1}
which may be beyond the reach 
of the LHC or a relatively light but highly compressed 
spectrum~\cite{compressed-SUSY-1} 
with soft decay products that escape detection. 
MSSM Higgs searches are complementary to searches for
colored scalars and electroweak gauginos.

The production modes of the neutral MSSM Higgs bosons are similar to those
of the SM Higgs boson with the most significant contributions coming from 
gluon fusion, weak boson fusion, and associated production with heavy
quarks. The associated production with 
one $b$ quark~\cite{Choudhury,Huang,Scott2002,Cao,Dawson:2007ur} 
or two $b$ quarks~\cite{Dicus1,hbbmm,Plumper,Dittmaier,Dawson:2003kb}  
can be enhanced by a large $\tan\beta$ and can produce a large cross section 
for even a heavy pseudoscalar Higgs. These $\tan\beta$ enhanced 
production modes with Higgs decaying into 
bottom quark pairs~\cite{hbbb1,hbbb2} and muon pairs~\cite{hbmm}, 
as well as Higgs decaying into tau pairs~\cite{hbll} provide promising
channels to discover the neutral Higgs bosons of the MSSM.
The best tau pair discovery channel for Higgs bosons 
has one tau decaying into a tau-jet ($\pi, \rho$ or $a_1$) 
and another decaying into a light charged lepton ($\ell = e$ or $\mu$). 
ATLAS and CMS groups have also looked into these channels and set put 
limits on the masses and $\sigma \times Br$ of the neutral MSSM Higgs 
bosons \cite{ATLAS-MSSM-Higgs,CMS-MSSM-Higgs}.

The inclusive tau pair discovery 
channel~\cite{Kunszt,Richter-Was,Carena:2005ek,Carena:2013qia} 
($pp \to \phi^0 \to \tau^+\tau^- +X, \phi^0 = h^0, H^0, A^0$) 
has been found to be very promising for the the search of 
neutral MSSM Higgs boson at the LHC.
In this article we study the associated production of neutral MSSM
Higgs bosons with a single $b$ quark with the Higgs decaying
subsequently into $\tau$ pairs followed by the decay of
$\tau$'s into leptons ($e^\pm\mu^\mp$). 
Although the decay rate is lower compared to the $\tau$-jet +
lepton channel, this channel does not suffer from the difficulties 
and uncertainties to tag a $\tau$-jet and provides an alternative 
with a cleaner signal containing two leptons. In the following sections
we study the Higgs signal with SUSY correction as well as the physics 
background, describe the acceptance cuts we employ and exhibit the LHC 
discovery potential of the MSSM neutral Higgs bosons
in this $be\mu$ channel.

\section{The Higgs Signal with Leptons}

The signal we consider is the associated production of a neutral MSSM
Higgs boson with a single $b$ quark followed by the decay of the Higgs
into a $\tau^+\tau^-$ pair and taus decaying 
into opposite sign different flavor leptons ($e^\pm\mu^\mp$) and neutrinos, i.e.
\begin{eqnarray*}
bg \to b\phi^0 \to b \tau^+ \tau^- \to b e^\pm \mu^\mp + \MET
\end{eqnarray*}
where $\phi^0 = h^0, H^0, A^0$. This search channel is complementary 
to the other important final state with a larger branching 
fraction $b\tau^+\tau^- \to b j_\tau \ell +\MET$.
Furthermore, this $be\mu$ discovery channel offers a cleaner signal
without the uncertainties involved with tau tagging and
avoids the physics background from $Z$ decay and the QCD background 
involving jets.

We calculate the cross section of the Higgs signal 
in $pp$ collisions $\sigma(pp \to bA^0 \to b\tau^+\tau^- +X)$ 
with a Breit-Wigner resonance via $bg \to bA^0 \to b\tau^+\tau^-$.
In our parton level calculations we use the leading order (LO) parton 
distribution function of CTEQ6L1 \cite{cteq6l1}.  
To include the next-to-leading order (NLO)
effects we choose both the factorization and renormalization scales to be
$M_\phi/4$~\cite{Willenbrock,Boos,Dawson:2004sh} with a K factor to be one.

The leading SM QCD and SUSY corrections to the bottom quark Yukawa
coupling can be calculated by using an effective Lagrangian 
approach~\cite{Carena:1999py}. 
For large $\tbeta$, the effective Lagrangian expressed in terms of the 
physical Higgs fields is given by
\begin{equation}
\Lagr =  \frac{(\bar{m}_b/v)}{1+\Delta_b} 
\left[ 
\left( \frac{\salpha}{\cbeta} - \Delta_b \frac{\calpha}{\sbeta}
\right) \bar{b}b h^0 
- \left( \frac{\calpha}{\cbeta} + \Delta_b \frac{\salpha}{\sbeta}
\right) \bar{b}b H^0
+ i\tbeta \bar{b} \gamma_5 b A^0 \right]
\end{equation}
where $\bar{m}_b$ denotes the running bottom quark mass including SM
QCD corrections which we evaluate with $m_b$(pole) = 4.7 GeV,
$v$ is the Higgs vacuum expectation value (VEV), and $\alpha$ is 
the mixing angle between the CP-even states $h^0$ and $H^0$. 
The function $\Delta_b$ includes loop suppressed threshold corrections 
from sbottom-gluino and stop-higgsino loops. 
In the large $M_{\rm SUSY}$ and $\tan\beta$ limit
$\Delta_b$ reads~\cite{Hall:1993gn,Carena:1994bv}
\begin{equation}
\Delta_b = \frac{2\alpha_s}{3\pi} \mgluino \mu \tbeta 
           \times I(\msbottom1,\msbottom2,\mgluino) 
+ \frac{\alpha_t}{4\pi} A_t \mu \tbeta \times I(\mstop1,\mstop2,\mu)
\end{equation}
where the auxiliary function $I$ is given by
\begin{equation}
I(a,b,c) = -\frac{1}{(a^2-b^2)(b^2-c^2)(c^2-a^2)} 
\left[ a^2b^2\log\frac{a^2}{b^2} + b^2c^2\log
\frac{b^2}{c^2} + c^2a^2 \log\frac{c^2}{a^2} \right].
\end{equation}
In our analysis of SUSY effects, we adopt the
conventions in Refs.~\cite{Dawson:2007ur,Dawson:2007wh}.
The branching width of the neutral Higgs bosons into the $b\bar{b}$
final state is also affected by these SUSY corrections which
indirectly affect the branching width into the $\tau^+\tau^-$ final
state as well. In the large $\tbeta$ limit, these branching ratios
\cite{Carena:2005ek} are approximately given by
\begin{eqnarray}
Br(A^0 \rightarrow b\bar{b} ) &\simeq& \frac{9}{(1+\Delta_b)^2 + 9} \\
Br(A^0 \rightarrow \tau^+\tau^- ) &\simeq&
\frac{(1+\Delta_b)^2}{(1+\Delta_b)^2 + 9} .
\end{eqnarray}
Therefore the cross section of our Higgs signal is approximately
\begin{eqnarray}
\sigma(bg \rightarrow bA^0 \rightarrow b\tau^+\tau^-)
&\simeq& \sigma_{SM} \times \frac{\tan^2\beta}{(1+\Delta_b)^2 + 9} 
 \simeq \frac{\sigma(\Delta_b = 0)}{(1+\Delta_b)^2 + 9}
\end{eqnarray}
which has only a mild dependence on $\Delta_b$. Depending on the sign
of $\mu$, which determines the sign of $\Delta_b$, these SUSY
corrections can enhance ($\mu < 0$) or suppress ($\mu > 0$) our signal. 
We study the neutral MSSM Higgs sector up to a TeV and assume 
all SUSY particles are heavy and above the Higgs sector. 
For $M_{\rm SUSY} = \msquark = \mgluino = A_t = 1 \sTeV$, $\mu = +200 \sGeV$ 
and $\tbeta = 10 \; (50)$ this corresponds to a $\sim 1 \; (4)\%$ 
drop in our signal cross section, for 
$M_{\rm SUSY} = \msquark = \mgluino = A_t = 2 \sTeV$, $\mu = +1 \sTeV$ 
and $\tbeta = 10 \; (50)$ we get a suppression of $\sim 2 \; (9)\%$. 
Since these effects are small for a large $M_{\rm SUSY}$, we neglect
them in the rest of our analysis.

\section{Higgs Mass Reconstruction}

The $\tau^+\tau^-$ decay mode of the Higgs generates large missing transverse momentum
due to the neutrinos in the final state which would normally make the mass
reconstruction difficult. But since the neutral Higgs bosons are 
much more massive than $\tau$'s ($m_\phi \gg m_\tau$), 
$\tau$'s produced in a Higgs decay 
are highly boosted, and their decay products --leptons and neutrinos
are almost collinear in the lab frame.
We exploit this kinematic feature and reconstruct the Higgs mass 
in the collinear approximation~\cite{Hagiwara:1989fn,Plehn:1999xi}. 
In the collinear limit, the decay product of each $\tau$ lepton 
can be identified by the fraction of energy it carries. 
Denoting these energy fractions with $x_1$ and $x_2$, 
the total missing transverse momentum can be expressed in terms
of the transverse lepton momenta as
\begin{eqnarray}
\vec{\slashed{p}}_T  = \left [ \frac{1}{x_1} -1 \right ] \vec{p}_T(\ell_1) + \left [ \frac{1}{x_2} -1 \right ] \vec{p}_T(\ell_2).
\end{eqnarray}
Given the measurements of the transverse momentum of charged leptons
and the missing transverse momentum, the above relation can be used
to determine the momenta of $\tau$'s:
\begin{eqnarray}
p^\mu(\tau_i) = \frac{p^\mu(\ell_i)}{x_i} \, , \quad i = 1, 2 \, .
\end{eqnarray}
Thus the Higgs mass can be reconstructed from the invariant mass of the $\tau$ pairs~\cite{Plehn:1999xi,Ellis:1987xu} as
\begin{eqnarray}
M_\phi = [p(\tau_1)+p(\tau_2)]^2
 = \left[ \frac{p(\ell_1)}{x_1} +\frac{p(\ell_2)}{x_2} \right]^2 \, .
\end{eqnarray}
For a physical solution, $x_{1,2}$ should be between 0 and 1. 
This physical solution requirement is one of the most effective cuts 
to reduce the SM background. To avoid large determinants that would
also imply large uncertainties in the solution we require the leptons 
not to be back to back in the transverse plane 
($\Delta \phi_T(e,\mu)<175^{\circ}$)~\cite{ATLAS-htata,CMS-htata}. 
We also require the leptons not to be parallel in the transverse plane 
in order to reduce the Drell-Yan and $t\bar{t}$ backgrounds 
($\Delta \phi_T(e,\mu)>5^{\circ}$)~\cite{ATLAS-htata}. 

 
\begin{figure}[htb]
\begin{center}
\includegraphics[scale=0.80]{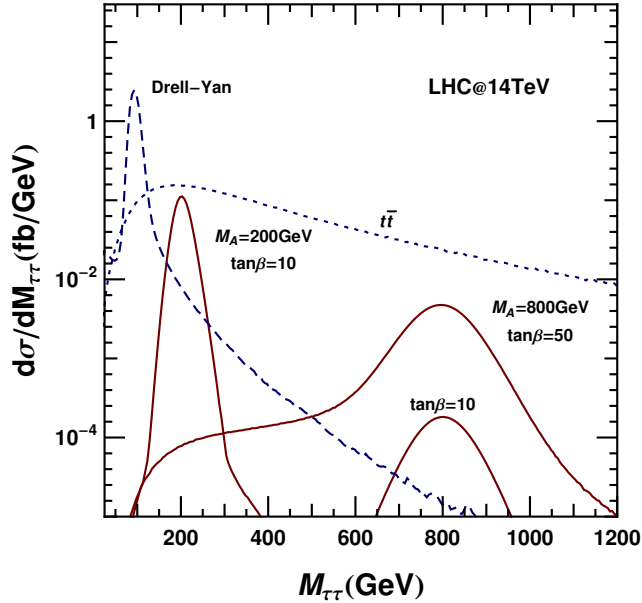}
\caption{\label{fig:mass}
The invariant-mass distribution, 
$d\sigma/dM_{\tau\tau}(pp \to b\tau^+\tau^- \to b e^\pm \mu^\mp +\MET +X)$, 
for the Higgs signal (solid red) from 
$bg \to bA^0 \to \tau^+ \tau^- \to b e^\pm \mu^\mp +X$ 
with $M_A = 200$ GeV and $\tan\beta = 10$ 
as well as $M_A = 800$ GeV for $\tan\beta = 10$ and $\tan\beta = 50$. 
Also shown is the physics background from the Drell-Yan process
$bg \to b \tau^+ \tau^- \to b e^\pm \mu^\mp +X$ (dashed blue) and 
from the $t\bar{t}$ process (dotted blue).
}
\end{center}
\end{figure}

In Figure~\ref{fig:mass} we present the invariant mass distribution of
the tau pairs for the Higgs signal $pp \to bA^0 \to b\tau^+\tau^- +X$ 
via $bg \to bA^0$, as well as the SM backgrounds due to Drell-Yan
production and top pair production. 
In this figure we have applied all acceptance cuts 
discussed in the next two sections
except the requirement on invariant mass. 


\section{The Physics Background}

The physics background consists of the following processes
\begin{eqnarray}
bZ/\gamma^* & \to & b \tau^+\tau^-  \to b e^\pm \mu^\mp + \MET \nonumber \\
jZ/\gamma^* & \to & j \tau^+ \tau^- \to j_b e^\pm \mu^\mp + \MET \nonumber \\
t\bar{t} & \to & \slashed{b} b e^\pm \mu^\mp + \MET \\
tW & \to & b e^\pm \mu^\mp + \MET \nonumber \\
jWW & \to & j_b e^\pm \mu^\mp + \MET \nonumber
\end{eqnarray}
where $j = q, g$ represents a light jet. 
We use the notation of $j_b$ to denote a light jet misidentified
as a $b$-jet and $\slashed{b}$ to denote a $b$-jet that escapes
detection. At low mass, due to the large $Z$ mass peak 
the dominant background is the Drell-Yan
process $pp \to b Z/\gamma^* \to b \tau^+ \tau^- + X$ and 
$pp \to j Z/\gamma^* \to b \tau^+ \tau^- + X$. 
At intermediate and high masses, Drell-Yan processes are suppressed as we
move away from the $Z$ pole and $t\bar{t}$ and $tW$ quickly become
dominant. The $jWW$ background is small due to the destructive 
interference between the Feynman diagrams that contribute to the same 
final states and due to the requirement of a light jet to be 
mistagged as a $b$-jet.

For the Drell-Yan processes, the different flavor leptons that we
require in the final state can only be produced through an initial
$\tau$ pair. But for the remaining background processes they
can be produced directly from $W$'s or indirectly by intermediate $\tau$'s. 
The branching ratio for leptonically decaying
$\tau$ ($\tau \to e \tilde{\nu}_e \nu_\tau / \mu \tilde{\nu}_\mu
\nu_\tau$'s) is about $17\%$. Hence each intermediate $\tau$
suppresses a channel approximately by the same amount. We calculate
all the contributions (0,1,2 intermediate $\tau$) except for the $jWW$
background for which we only consider $W$'s decaying directly into $e$ or $\mu$
since the cross section of this process is already quite small.

\section{Acceptance cuts}

To simulate the detector effects, we apply Gaussian smearing with the
energy measurement uncertainty parametrized by an energy dependent term and an
energy independent term added in quadrature as
\begin{eqnarray}
\frac{\Delta E}{E} = \frac{a}{\sqrt{E}} \oplus b
\end{eqnarray}
where we use $a = 60\%(25\%)$ and $b=3\%(1\%)$ for jets (leptons)
following the ATLAS and CMS TDR~\cite{ATLAS-TDR,CMS-TDR}. 
We assume a constant $b$-tagging efficiency
throughout the detector with the rate $\epsilon_b = 60\%$, and constant
mistagging rates of $c$-jets and light jets as $b$-jets with the rates
$\epsilon_c = 14\%$ an $\epsilon_j = 1\%$.

\begin{table}[!ht]
\renewcommand{\arraystretch}{1.4}
\centering
\begin{tabular}{|C{6.0cm}|C{6.0cm}|}
\hline
\multicolumn{2}{|c|}{Acceptance cuts (LL,HL)}  \\
\hline
$p_T(b) > (20,30) \sGeV$ & $|\eta(b,e,\mu)| < 2.5$ \\
$p_T(e,\mu) > (15,20) \sGeV$ & $\Delta R(b,e,\mu) > 0.4$  \\
$\MET > (20,40) \sGeV$ & $5 \degree < \Delta \phi_T(e,\mu) < 175 \degree$ \\
$|M_{\tau \tau}-M_A| < (0.15,0.20) \times M_A$ & $0 < x_{1,2} < 1$  \\
\hline
\end{tabular}
\caption{\label{cuts}
Acceptance cuts for low and high luminosity (LL,HL). We veto two jet events for which $p_T(b_1,b_2) > 20 \sGeV$ and $|\eta| < 4.5$.}
\end{table}

In order to account for the noisy detector environment due to pile-up,
we employ two sets of cuts specific for low and high
luminosity (LL,HL). We require exactly one high transverse momentum
$b$-tagged jet and two opposite sign different flavor leptons in the
event. The $b$-jet is required to have $p_T > 20 \sGeV$ (LL) or $p_T >
30 \sGeV$ (HL) and $|\eta| < 2.5$. To reduce the $t\bar{t}$ background
we veto two jet events with $p_T > 20 \sGeV$ and 
$|\eta| < 4.5$~\cite{CMS:2013aea}. 
We require both leptons to be isolated by imposing $\Delta R > 0.4$ and
to have $p_T > 15 \sGeV$ (LL) or $p_T > 20 \sGeV$ (HL). We apply a $20
\sGeV$ (LL) and $40 \sGeV$ (HL) cut on the missing transverse momentum
which we define as the negative sum of the transverse momenta of the
visible objects in the event. We finally require the reconstructed
Higgs mass to be within $15\%$ (LL) or $20\%$ (HL) of the pseudoscalar
Higgs mass $M_A$. A summary of the basic cuts we employed is displayed
in Table \ref{cuts}.

We use MadGraph~\cite{MadGraph5} to generate HELAS~\cite{HELAS} 
subroutines to compute the matrix elements for the tree level signal 
and background processes. We introduce the NLO corrections to 
the SM background processes as $K$ factors. We apply 
a $K$ factor 1.3 for the Drell-Yan 
processes~\cite{Drell-Yan-kfactor}, 
a $K$ factor of 2 for top pair 
production~\cite{ttbar-kfactor-1,ttbar-kfactor-2}, 
a $K$ factor of 1.58 for $tW$ production~\cite{tw-kfactor}, 
and a $K$ factor of 1 for $jWW$ background. Cross sections 
and signal significance for benchmark points are displayed in Table
\ref{benchmarks}. 

\begin{figure*}[htb]
\begin{center}
\includegraphics[scale=0.80]{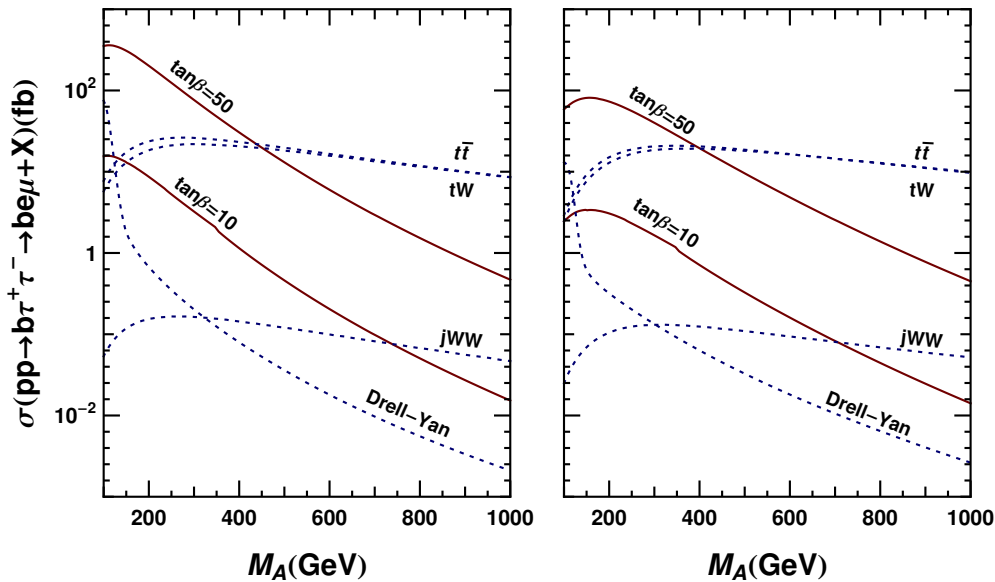}
\caption{\label{fig:sigma}
Signal and background cross sections at $14 \sTeV$ after all the cuts
(Table \ref{cuts}) are applied. Left panel shows the cross sections
with low luminosity (LL) cuts, right panel shows the cross sections
with high luminosity (HL) cuts. Total cross section for the Higgs
signal with $\tan\beta = 10,50$ (solid red) and the SM background 
processes ($t\bar{t}$, $tW$, $jWW$, Drell-Yan) (dashed blue) 
are displayed.}
\end{center}
\end{figure*}

Figure~\ref{fig:sigma} shows the signal and background cross sections with
$\sqrt{s} = 14$ TeV and acceptance cuts for low luminosity (LL) and 
high luminosity (HL) as a function of the pseudoscalar Higgs mass $M_A$.
The signal is shown for 
$\tan\beta = 10$ and 50, with a common mass for scalar quarks, scalar 
leptons, gluino, and the $\mu$ parameter from the Higgs term in the 
superpotential, 
$m_{\tilde{q}} = m_{\tilde{g}} = m_{\tilde{\ell}} = \mu = 1$ TeV. 
All tagging efficiencies and $K$ factors discussed above are included. 

\begin{table}[!ht]
\renewcommand{\arraystretch}{1.4}
\centering
\begin{tabular}{|C{4.3cm}|C{1.17cm}C{1.17cm}C{1.17cm}C{1.17cm}|}
\hline
$M_A (\GeV)$ & 100 & 200 & 400 & 800 \\
\hline
$\sigma(\textrm{signal}) \quad [\tan \beta = 10]$ & 15.85 & 8.847 & 1.176 & 0.052 \\
$\sigma(\textrm{signal}) \quad [\tan \beta = 50]$ & 344.0 & 196.8 & 29.82 & 1.496 \\
\hline
$\sigma(\textrm{Drell-Yan})$ & 57.40 & 0.517 & 0.061 & 0.004 \\
$\sigma(t \bar{t})$ & 2.867 & 9.158 & 10.25 & 5.792 \\
$\sigma(tW)$ & 4.884 & 14.96 & 14.71 & 7.521 \\
$\sigma(jWW)$ & 0.054 & 0.153 & 0.144 & 0.067 \\
\hline
$N_{sig} \quad [\tan \beta = 10]$ & 8.513 & 6.745 & 0.959 & 0.059 \\
$N_{sig} \quad [\tan \beta = 50]$ & 90.64 & 69.65 & 19.02 & 1.637 \\
\hline
\end{tabular}
\caption{\label{benchmarks}
Signal and SM background cross sections in femtobarns and signal
significance, i.e. $\sigma_S/\sqrt{\sigma_S+\sigma_B}$
for $\sqrt{s} = 14 \sTeV$ and $L = 30 \sifb$. $K$ factors are included 
in the signal significance values.}
\end{table}

We use $\hat{s}_{min}^{1/2}$ to further reduce the Standard Model
backgrounds, which is a global and fully inclusive variable designed
to determine the mass scale involved in a scattering event with
missing energy \cite{shatmin}. It is defined as
\begin{eqnarray}
\hat{s}_{min}^{1/2} = \sqrt{E^2 - P_z^2} + \sqrt{\slashed{E}_T^2 + M^2_{inv}}
\end{eqnarray}
where $M_{inv}$ is the total mass of all invisible particles produced
in an event. In our case the invisible particles are neutrinos hence
we set $M_{inv} = 0$.

\begin{figure}[htb]
\begin{center}
\includegraphics[scale=0.80]{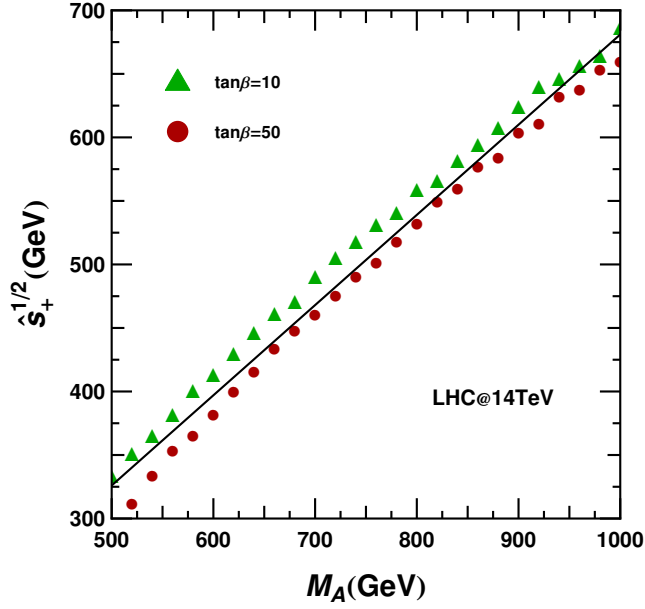}
\caption{\label{fig:smin}
Optimal $\hat{s}_{min}^{1/2}$ cut as a function of the
pseudoscalar Higgs mass $M_A$ for $\tan\beta=10$ (green triangles) and $\tan\beta=50$ (red circles). The best fit is approximately 
given by $\hat{s}_{+}^{1/2} = 0.7 \times M_A$.}
\end{center}
\end{figure}

For the Higgs signal, the mass of the Higgs particle determines the
minimum center of mass energy of the process since the Higgs is mostly
on-shell. Similarly for the background, intermediate on-shell
particles determine the mass scale. For the main $t\bar{t}$ background
the mass scale is $2m_t$. So this variable is effective in reducing
the $t\bar{t}$ and $tW$ backgrounds in the high mass region where $M_A
> 2 m_t$. What we actually get from $\hat{s}_{min}^{1/2}$ is not
exactly the mass of the intermediate particles but an event by event
lower bound of the center of mass energy of the hard interaction.
Therefore we expect this variable to be effective well above 
the $2m_t$ threshold. To optimize our cut, we determine the 
$\hat{s}_{+}^{1/2}$ value for which the cut
\begin{eqnarray}
\hat{s}_{min}^{1/2} > \hat{s}_{+}^{1/2}
\end{eqnarray}
maximizes the signal significance,
i.e. $\sigma_S/\sqrt{\sigma_S+\sigma_B}$. Since the mass scale for the
Higgs signal changes with $M_A$, the optimum cut $\hat{s}_{+}^{1/2}$
depends on $m_A$ as well. To determine its $M_A$ dependence we do a scan
over $M_A$ in the range [500 GeV, 1000 GeV] for $\tan\beta=10,50$ and
compute the optimum $\hat{s}_{min}^{1/2}$ cut.
We display the result of this scan in Figure~\ref{fig:smin}.

We observe that the shape of the $\hat{s}_{min}^{1/2}$ distribution 
for the SM background does not change significantly with our Higgs mass window
cut, but for the Higgs signal it shifts towards higher values with
increasing Higgs mass while broadening due to more missing energy
carried away by neutrinos. This results in an almost linear relation between
the optimum $\hat{s}_{+}^{1/2}$ cut and $m_A$ which we determine 
to be $\hat{s}_{+}^{1/2} = 0.71 \times  M_A -29 \sGeV$. As can be seen 
from Figure~\ref{fig:smin} the optimum value
has a small $\tan\beta$ dependence as well. In the rest of our analysis
we use $\hat{s}_{+}^{1/2} = 0.7 \times M_A$ for simplicity.

\section{The Discovery Potential at the LHC}

\begin{figure}[!htb]
\begin{center}
\includegraphics[scale=0.83]{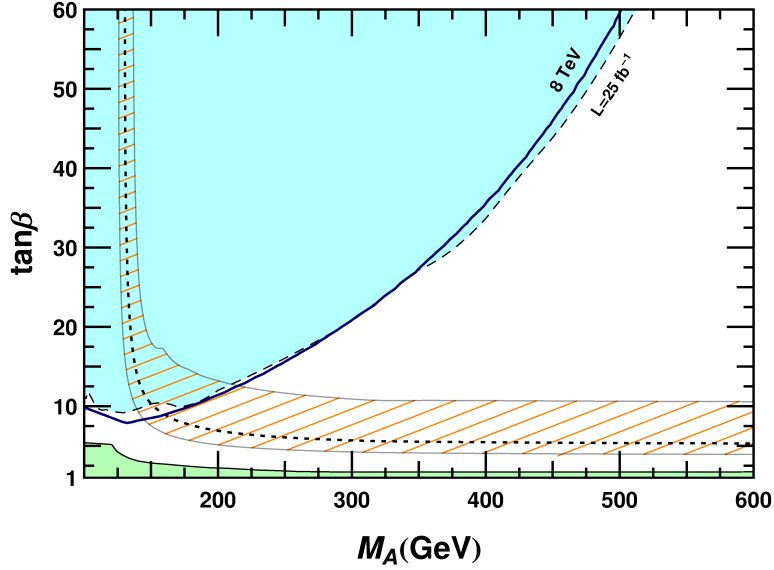}
\caption{\label{fig:reach8}
The $5\sigma$ discovery contour at the LHC with 
$\sqrt{s} = 8 \sTeV$ and a luminosity of $L = 25 \sifb$. 
The discovery region is the part of the parameter space above the contour.
Also shown are (a) the region excluded by LEP II (green, lower
shaded), (b) the region excluded by LHC Higgs searches (cyan, upper
shaded), and (c) the region with a favored light Higgs mass of
123 GeV $\le m_h\le$ 129 GeV (orange, hatched) and the central value of 126 GeV (dotted).}
\end{center}
\end{figure}

\begin{figure}[!htb]
\begin{center}
\includegraphics[scale=0.83]{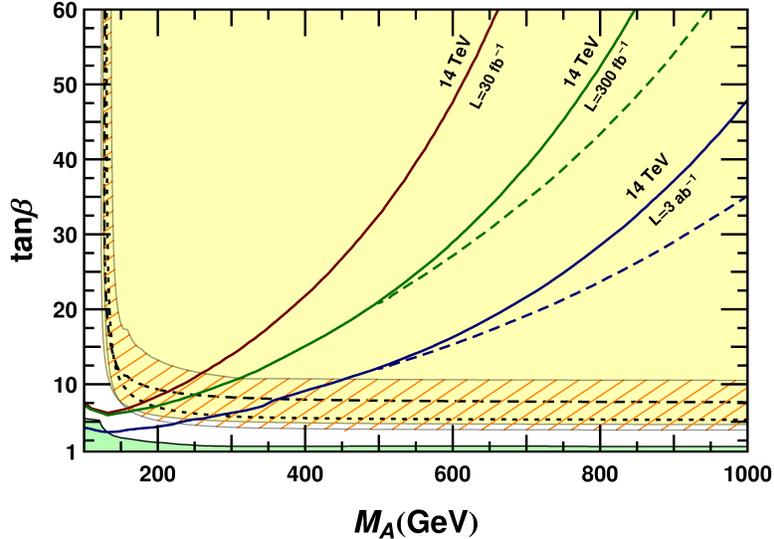}
\caption{\label{fig:reach14}
The $5\sigma$ discovery contours at the LHC with $\sqrt{s} = 14 \sTeV$ 
and a luminosity of $L = 30, 300, 3000 \sifb$.
The dashed lines show the improvement obtained with the aid of 
the $\hat{s}_{min}^{1/2}$ cut.
The discovery region is the part of the parameter space above the contours.
Also shown are (a) the region excluded by LEP II (green, lower
shaded), and (b) the region with a favored light Higgs mass of
123 GeV $\le m_h\le$ 129 GeV (orange and yellow, hatched and shaded) and the central value of 126 GeV (dotted and dashed).}
\end{center}
\end{figure}

To calculate the LHC reach, we scan the $(M_A,\tan\beta)$ plane and
display the discovery contours for $\sqrt{s} = 8 \sTeV$ 
with an integrated luminosity $L = 25 \sifb$ 
as well as $\sqrt{s} = 14 \sTeV$ with 
integrated luminosities $L = 30 \sifb$, $300 \sifb$, $3 \siab$ in
Figures \ref{fig:reach8} and \ref{fig:reach14}. 
In addition, we also show the improvement with the addition of the
$\hat{s}_{min}^{1/2}$ cut.

We define the signal to be observable 
if the lower limit on the signal plus background is larger than 
the corresponding upper limit on the background \cite{HGG,Brown}, namely,
\begin{eqnarray}
L (\sigma_S+\sigma_B) - N\sqrt{ L(\sigma_S+\sigma_B) } > 
L \sigma_B +N \sqrt{ L\sigma_B }\,\,,
\end{eqnarray}
which corresponds to
\begin{eqnarray}
\sigma_S > \frac{N^2}{L} \left[ 1+2\sqrt{L\sigma_B}/N \right]\,\,.
\end{eqnarray}
Here $L$ is the integrated luminosity, 
$\sigma_S$ is the signal cross section, 
and $\sigma_B$ is the background cross section.  
Both cross sections are taken to be 
within a bin of width $\pm\Delta M_{\tau\tau}$ centered at $M_A$. 
In this convention, $N = 2.5$  corresponds to a 5$\sigma$ signal.

For $\tan\beta \agt 10$, $M_A$ and $M_H$ are almost degenerate 
when $M_A \agt$ 125 GeV, while $M_A$ and $M_h$ are very close to 
each other for $M_A \alt$ 125 GeV \cite{Higgsmass1,Higgsmass2}.
Therefore, when computing the realistic discovery reach, 
we add the cross sections of the $A^0$ and the $h^0$ for $M_A < 125$ GeV 
and those of the $A^0$ and the $H^0$ for $M_A \ge 125$ GeV 
\cite{Kao:1995gx}. 

We use FeynHiggs~\cite{FeynHiggs} to calculate the light Higgs mass at 
two loop level \cite{Ellis:1990nz,Heinemeyer:1998np,Degrassi:2002fi}. To cope with the 
remaining theory uncertainty in the light Higgs mass which
is about 2-3 GeV \cite{Degrassi:2002fi}, we define a favored light 
Higgs mass band (for a 126 GeV light Higgs) to be the range 
123 GeV $\le m_h\le$ 129 GeV.

Figure~\ref{fig:reach8} shows the 5$\sigma$ discovery contour in the
($M_A,\tan\beta$) plane for the neutral MSSM Higgs bosons at the LHC 
with $\sqrt{s} =$ 8 TeV and $L = 25$ fb$^{-1}$. 
Also shown is the parameter region excluded by LEP II~\cite{LEP2}. 
In addition, we present the favored region of a light Higgs boson 
(123 GeV $\le m_h\le$ 129 GeV) for 
$M_{\rm SUSY} = m_{\tilde{q}} = m_{\tilde{g}}
 = m_{\tilde{\ell}} = \mu = 1$ TeV and  
$X_t = A_t -\mu\cot\beta = 2$ TeV, 
where $A_t$ is the trilinear coupling for scalar top.

Figure~\ref{fig:reach14} shows the 5$\sigma$ discovery contours for 
the MSSM Higgs bosons at the LHC with $\sqrt{s} =$ 14 TeV with 
$L = 30, 300$ and $3000$ fb$^{-1}$. 
We display again the regions with a favored light Higgs mass 
(123 GeV $\le m_h\le$ 129 GeV) for 
$M_{\rm SUSY} = 1 \sTeV (2 \sTeV)$ and $X_t = 2 \sTeV (\sqrt{6}M_{\rm SUSY})$.
We find that the discovery contour even dips below $\tan\beta = 10$ for 
$100$ GeV $< M_A < 300 - 400$ GeV depending on luminosity. 
Below $\tan\beta = 10$ our approximation of mass degeneracy of MSSM 
Higgs bosons breaks down; therefore we include only one Higgs boson
$(A^0)$ in our calculations to simplify the numerical analysis.
For $M_A,M_H \agt 400$ GeV the Higgs cross section becomes kinematically 
suppressed while for lower masses ($M_A \alt 300$ GeV), the Higgs
cross section is reasonably large. Therefore, for $M_A \alt
300$ GeV even the CP-odd pseudoscalar alone can lead to an observable
signal with $5 < \tan\beta < 10$. High mass regions with $M_A, M_H
\agt 400$ can be probed if $\tbeta$ is large. 
Specifically for $M_A = 1 \sTeV$ and $\tbeta = 60$, 
MSSM neutral Higgs bosons can provide a $5\sigma$ discovery signal 
with an integrated luminosity of $L \simeq 300 \sifb$.

\section{Conclusions}

We have studied the production of neutral MSSM Higgs bosons at the LHC
associated with a single $b$ quark followed by Higgs decay into tau
pairs and tau leptons decaying to electron-muon
pairs. This production channel is enhanced for large $\tan\beta$ and 
this specific final state offers a clean signal albeit a
smaller branching ratio compared to the more promising 
tau pair discovery channel with $b\tau^+\tau^- \to b j_\tau \ell +\MET$.
The $b e \mu$ channel does not require tau jet tagging hence eliminates the 
uncertainties involved with it, and the physics background for our signal 
from $Z$ decay and the QCD backgrounds containing light jets are more suppressed.

Motivated with the latest non-observation of super partners, we have considered
a heavy SUSY spectrum with squarks, sleptons and the gluino above the Higgs sector. 
After all the cuts are applied, the Higgs signal cross section is about $1.5 \sfb$ for
$M_A = 800 \sGeV$ and $\tbeta = 50$ at the LHC running at 14 TeV center of mass energy.
We have calculated the relevant background
processes which are Drell-Yan, $t\bar{t}$, $tW$ and $jWW$ productions
with full spin correlation. The Drell-Yan background is dominant at low mass and 
$t \bar{t} / tW$ backgrounds are dominant at high mass regions. Our calculation 
shows that the discovery contour for an integrated luminosity of $L = 300 \sifb$ 
extends to $M_A = 800 \sGeV$ for $\tan\beta=50$ and up to almost $M_A = 1 \sTeV$ 
for $\tan\beta=60$ with the help of the $\hat{s}_{min}^{1/2}$ variable.

\begin{acknowledgments}

We are grateful to Howie Baer, Sally Dawson and Phil Gutierrez for useful discussions. 
The computing for this project was performed at the OU Supercomputing
Center for Education \& Research (OSCER) at the University of Oklahoma.
This research was supported in part by the U.S. Department of Energy 
under Grant No.~DE-FG02-13ER41979. 

\end{acknowledgments}


\end{document}